\documentstyle[12pt]{article}
\def\bbox{{\,\lower0.9pt\vbox{\hrule \hbox{\vrule height 0.2 cm
\hskip 0.2 cm
\vrule  height 0.2 cm}\hrule}\,}}

\newcommand{\beq}{\begin{equation}}
\newcommand{\eeq}{\end{equation}}
\newcommand{\bea}{\begin{eqnarray}}
\newcommand{\eea}{\end{eqnarray}}

\newcommand{\req}[1]{Eq.(\ref{#1})}

\begin{document}
\setlength{\unitlength}{1mm}
\title{{\hfill {\small } } \vspace*{2cm} \\
Thermodynamics and Statistical Mechanics of Induced Liouville
Gravity}
\author{\\
V. Frolov${}^1$, D. Fursaev${}^2$, J. Gegenberg${}^3$ and G. Kunstatter${}^4$
 \date{January, 1999}}
\maketitle
\noindent  {
$^{1}${ \em
Theoretical Physics Institute, Department of Physics, \ University of
Alberta, \\ Edmonton, Canada T6G 2J1}
\\ $^{2}${\em Joint Institute for
Nuclear Research,\\
Bogoliubov Laboratory of Theoretical Physics,\\
 141 980 Dubna, Russia}
\\${}^3$ {\em Dept. of Math. and Statistics,\\
University of New Brunswick, Fredericton, NB E3B 5A3 }
\\${}^4${\em
 Physics Department and Winnipeg Institute of Theoretical Physics\\
University of Winnipeg, Winnipeg, MB R3B 2E9}
\\
e-mails: frolov@phys.ualberta.ca, fursaev@thsun1.jinr.ru,
gabor@theory.uwinnipeg.ca, \\
lenin@math.unb.ca
}
\bigskip

\begin{abstract}
In this paper we describe a Liouville
gravity which is induced by a set of quantum fields
(constituents) and represents a two-dimensional analog of
Sakharov's induced gravity.
The important feature of the considered theory is
the presence of massless constituents which are responsible
for the appearance of the induced Liouville field.
The role of the massive constituents is only to induce
the cosmological constant.
We consider the instanton solutions of the Euclidean Liouville gravity
with negative and zero cosmological constants,
some instantons
being interpreted as
two-dimensional anti-de Sitter $AdS_2$ black holes.
We study thermodynamics of all the solutions
and conclude that their entropy
is completely determined by the
statistical-mechanical entropy of the massless constituents.
This shows, in particular, that the constituents  of
the induced gravity
are the true degrees of freedom of $AdS_2$ black holes.
Special attention is
also paid to
the induced Liouville gravity with zero cosmological constant on a
torus.
We demonstrate the equivalence of its thermodynamics to
the thermodynamics of BTZ black holes and comment on
computations of the BTZ black hole entropy.
\end{abstract}

\bigskip
%\vspace{3cm}

\baselineskip=.6cm

\newpage

\section{Introduction}
\setcounter{equation}0

The  microscopic explanation of the
Bekenstein-Hawking entropy \cite{Bekenstein:72},\cite{Hawking:75}
of black holes is
one of the most intriguing problems of theoretical physics.
Although there are several approaches to its resolution
(for a review of some of them
see \cite{FF:98a},\cite{Peet}) this problem is a subject of
 intensive study and continues to inspire  new ideas.

In particular, it was realized recently that at least some black holes
may be macroscopically equivalent to two-dimensional systems described
by a Liouville theory. The statistical-mechanical entropy of these
systems can be computed by means of a conformal field theory and
coincides with the black hole entropy.

This observation was first made for extremal black holes \cite{StVa}
and was based on dualities in string models.
However, in the most explicit and simple form it appeared
in the work by
Strominger \cite{Strominger}
concerning BTZ black holes \cite{BTZ} (for a detailed analysis
of these computations with the large list of references see
\cite{Carlip:98a}).
The BTZ black holes
have the same thermodynamic characteristics as a dual Liouville
theory defined at asymptotic infinity.
The reason for such a relation arises from a  specific property of
three-dimensional gravity with  negative cosmological constant.
It is equivalent to a Chern-Simons theory which
has only boundary degrees of freedom that are  described by a Liouville
theory \cite{BH},\cite{CHD}.

A Liouville theory which is dual to the  BTZ black hole can be also defined
at the black hole horizon \cite{Carlip:95}. Moreover,
close to the horizon one can find a Liouville-like
description of black holes in an arbitrary dimension
\cite{Carlip:98b},\cite{Solodukhin:98}. This description becomes
possible because the gravitational action for spherically symmetric
metrics is the action for a two-dimensional dilatonic gravity.
In the region near the horizon this gravity is equivalent to a
Liouville theory \cite{Solodukhin:98} and one can use the conformal
theory to calculate the entropy.

These results may have the following interpretation \cite{Martinec}.
On the level of
thermodynamics black holes are equivalent to a Liouville
field theory. The Liouville field, which is purely classical, is a
collective excitation of some quantum constituents which are described
by 2D conformal field theory. To put it in another way, the Liouville
theory is an effective theory of the constituents. The constituents are
those microscopic degrees of freedom which explain the thermodynamic
entropy related to the Liouville field and thus reproduce the black
hole entropy in a statistical-mechanical way.

Remarkably, this mechanism is basically the same as the
mechanism
of the generation of the black hole entropy in
Sakharov's induced gravity
\cite{Jacobson:94}--\cite{FF:98b}.
According to Sakharov's idea \cite{Sakharov:68},
the gravitational field is a collective excitation of the
matter constituents and the Einstein action is
the low-energy effective action of the constituents.
The equations for the metric $g_{\mu\nu}$ are
\begin{equation}\label{i1}
\langle \hat{T}_{\mu\nu}(g) \rangle=0~~~.
\end{equation}
Here $\hat{T}_{\mu\nu}$ is the stress-energy tensor
of constituent fields on the background with the metric
$g_{\mu\nu}$ and its average is taken in some
quantum state.

It should be noted that in Sakharov's induced gravity
the microscopic states of a black hole are related to the constituents
which live on the physical space-time.  In the Liouville description of
black holes the microscopic degrees of freedom live
on a dual two-dimensional space-time. For this reason the statistical
origin of
the Bekenstein-Hawking entropy in the two approaches is
different.

Nevertheless, the similarity of  both approaches
suggests that they may be connected.
To see whether there is
any connection one must first
understand better how  Sakharov's mechanism of induced gravity
works in the case of the Liouville theory. Studying
induced Liouville gravity is the subject of this paper.

\bigskip

The important feature of the considered models of
two-dimensional induced gravity
is the presence of massless constituents which induce the Liouville
action.  The massive constituents serve only to induce a finite
cosmological constant.  Thus, as distinct from higher dimensions, all
the dynamics in the Liouville induced gravity is due to massless
fields.
This fact has a crucial consequence for the statistical interpretation
of the thermodynamics  of the instanton solutions of the theory.

\bigskip

We consider models with negative and zero cosmological constants.
The theory with negative cosmological constant is
a sort of gravity on anti-de Sitter
$AdS_2$ space-time. Studying this theory is
motivated by various reasons, one of which is its relation to
$AdS_2$ string theories arising as near-horizon limits
of different four and  five-dimensional black holes
(for a recent discussion of $AdS_2$ gravity in this context
see \cite{St:98}).
The Liouville induced gravity with vanishing
cosmological constant is interesting because it is that theory which
is thermodynamically equivalent to BTZ black holes.

The paper is organized as follows. In Section 2 we use
the Sakharov's mechanism
to induce Liouville gravity by massless and massive
quantum constituent fields. The considered quantum models are free from
ultraviolet divergences and, as a result, the induced theory
is finite and well-defined.
All the solutions of Liouville gravity with the negative
cosmological constant are locally $AdS_2$.
In the Euclidean theory they can be of three types:  elliptic,
hyperbolic and parabolic. We remind the reader the form of these
solutions in Section 3. The instantons of the elliptic type can be
interpreted as $AdS_2$ black holes. Thermodynamics and
statistical mechanics of these black holes is studied in Section 4.
We demonstrate that in the physical processes the changes of
the entropy of
a $AdS_2$ black hole coincide with the corresponding
changes of the
entanglement (statistical-mechanical)
entropy of the massless constituents. Therefore, the
constituents of the induced gravity are the true internal degrees of
freedom of $AdS_2$ black hole.
In Section 5 we comment on thermodynamics and statistical mechanics
the hyperbolic and parabolic
solutions.
Finally, in Section 6 we consider induced
Liouville gravity with vanishing cosmological constant on a torus
and demonstrate the equivalence of its thermodynamics to
the thermodynamics of the BTZ black hole under
identifying the central charges of both theories.
A brief discussion is given in Section 7.

\section{Induced Liouville gravity}
\setcounter{equation}0

Induced Liouville gravity (ILG) can be constructed from
models with different constituent field species, similar
to the construction of induced Einstein gravity
\cite{FFZ:97}-\cite{FF:98b}.
To illustrate the idea we consider here the
simplest model.  It
consists of noninteracting scalar and spinor fields, with some of the scalar
fields being massless.  The numbers of massive and massless scalars are
$N_s$ and $N$, respectively, and the number of massive spinor fields is
$N_d$. It is assumed that some massive scalars are non-minimally
coupled and the corresponding constants are denoted as $\xi_s$. The
effective gravitational action of the fields propagating on a
background with the metric $g_{\mu\nu}$ is
\begin{equation}\label{1.1}
\Gamma=\sum_s W_s+\sum_d W_d+NW_s^0~~~,
\end{equation}
\begin{equation}\label{1.2a}
W_s=\frac 12\log\det\left(-\nabla^2+\xi_s R+
m_s^2\right)~~,
~~W_d=-\log\det\left(\gamma^\mu\nabla_\mu+m_d\right)~~,
\end{equation}
\begin{equation}\label{1.2b}
W_s^0=\frac 12\log\det\left(-\nabla^2\right)~~~,
\end{equation}
Here $R$ is the scalar curvature of the background.
It is not difficult to show that $\Gamma$ is free
from ultraviolet divergences if the following
constraints are satisfied
\begin{equation}\label{1.3}
N_s+N-2N_d=0~~,~~\sum_s m_s^2-2\sum_d m_d^2=0~~,
\end{equation}
\begin{equation}\label{1.4}
N_s+N+N_d-6\sum_s\xi_s=0~~.
\end{equation}
Constraints (\ref{1.3}) ensure the finiteness
of the induced cosmological constant, while condition (\ref{1.4})
guarantees the finiteness of the induced Newton constant.

\bigskip

Suppose now that masses $m_i$
have the order of magnitude of a typical mass $M$.
The low-energy limit of the theory
is realized when the curvature
of the background geometry
is small compared to $M^2$.
In this limit contributions of the massive constituents to
the induced action can be expanded in powers of the curvature.
On the other hand, the contributions of the
massless constituents, $W_s^0$, can be calculated exactly.

It is convenient
to represent the induced action in the form
\begin{equation}\label{1.11}
\Gamma=\Gamma^{\tiny{m}}+
NW
~~~,
\end{equation}
\begin{equation}\label{1.12}
\Gamma^{\tiny{\mbox{m}}}=\sum_s W_s+\sum_d
W_d+NW_{s,\tiny{\mbox{div}}}^0~~~,
\end{equation}
\begin{equation}\label{1.14}
W\equiv W_s^0-
W_{s,\tiny{\mbox{div}}}^0~~~.
\end{equation}
Here $W_{s,\tiny{\mbox{div}}}^0$ is the divergent part of the
action of massless fields, so that  $W$
is the "renormalized" action. Note that because
we are dealing with ultraviolet finite theories
the functionals $\Gamma^{\tiny{\mbox{m}}}$ are free
from the divergences. The divergences of the massive
fields in $\Gamma^{\tiny{\mbox{m}}}$ are canceled
by the term $NW_{s,\tiny{\mbox{div}}}^0$.

Let us consider fields given on a manifold ${\cal M}$
with  boundary $\partial {\cal M}$. The metric on
${\cal M}$ will be denoted $g_{\mu\nu}$. In the general case,
$g_{\mu\nu}$ can have Lorentzian or Euclidean signature.
In what follows we assume that ${\cal M}$ is a Euclidean manifold.
In the low energy limit of the theory the
action of the massive constituents $\Gamma^{\tiny{m}}$
can be expanded in powers of  the curvature. We keep only
the leading (cosmological) term
in this decomposition and approximate $\Gamma^{\tiny{m}}$
as
\begin{equation}\label{1.15}
\Gamma^{\tiny{m}}=\int_{\cal M} \sqrt{g} d^2x
\lambda~~~.
\end{equation}
Here the cosmological constant $\lambda$ is
\begin{equation}\label{1.16}
\lambda=-{1 \over 8\pi}\left(\sum_s m_s^2\ln m_s^2
-2\sum_d m_d^2\ln m_d^2\right)~~~.
\end{equation}
In what follows,
we consider only the models where the
cosmological constant is negative or zero.
The curvature corrections to expression (\ref{1.15})
are suppressed by powers $RM^{-2}$.

The above approximation is not applicable to the action
$W$ of massless
fields.  In fact, this functional is the
well-known Polyakov action \cite{Polyakov}
which can be computed exactly.
Consider the conformal map of $\cal M$
onto a space $\bar{\cal M}$ with the metric
$\bar{g}_{\mu\nu}=\exp(-2\sigma)g_{\mu\nu}$.
The actions on $\cal M$ and $\bar{\cal M}$
are related as
\begin{equation}\label{1.18}
W[g]=
W[\bar{g}]-
{1 \over 24\pi}\left[\int_{\cal M}\sqrt{g} d^2x
(R\sigma -(\nabla\sigma)^2)+
\int_{\partial {\cal M}} h^{1/2} dy (2K\sigma
+3n^\mu\sigma_{,\mu})\right]
~~~.
\end{equation}
Here  $n^\mu$ is
a unit vector normal to the boundary $\partial {\cal M}$
of $\cal M$; $K$ and $h$ are the extrinsic
curvature and the metric on $\partial {\cal M}$.
The functional $W[\bar{g}]$
is the effective action computed on the background
$\bar{\cal M}$ with the metric $\bar{g}$.
It is convenient to assume that $\bar{\cal M}$ is locally flat.

As follows from the above analysis,
the induced gravitational action
after subtracting a boundary term depending on $n^\mu\sigma_{,\mu}$
can be written in
the form:
\beq
\label{ac1}
\Gamma[g]= I_L[g,\phi]+ N W[\bar{g}]~~~,
\eeq
\beq
\label{l1}
I_L[g,\phi]=-{1 \over 8\pi}\int_{\cal M} \sqrt{g}d^2x
\left((\nabla \phi)^2+
{2 \over \gamma} R\phi +{\mu \over \gamma^2}\right)
-{1 \over 2\pi\gamma}\int_{\partial {\cal M}} h^{1/2} dy K\phi~~~.
\eeq
Here we put
\beq
\label{phi}
\phi={2 \over \gamma}\sigma~~~,
\eeq
\beq
\label{gamma}
\gamma=\sqrt{12 \over N}~~~,~~~\mu={96\pi \over N}|\lambda|~~~.
\eeq
The action $I_L$ can be also
represented as a functional on the flat space $\bar{\cal M}$
\beq
\label{l2}
I_L[\phi,g]=\bar{I}_L[\phi]=
-{1 \over 2\pi}\int_{\bar{\cal M}}
d^2y\left( (\bar{\nabla}\phi)^2+
{\mu \over \gamma^2} e^{\gamma \phi}\right)
-{\beta \over 2\pi\gamma}(\phi_+ -\phi_-)~~~.
\eeq
Here $\beta$ is the circumference length of the boundary, and
$\phi_+$, $\phi_-$
are the values of $\phi$ on the external and internal parts of the
boundary.
If the internal boundary is absent
$\phi_-=0$ in (\ref{l2}).

Up to the boundary term, $\bar{I}_L[\phi]$ is the canonical
Euclidean Liouville action \footnote{Strictly speaking,
$\bar{I}_L[\phi]$ differs from the standard definition by a
sign, see \cite{Seiberg}.}.
The Liouville theory is known from the last century as a theory
of negatively curved surfaces.
A review of some its properties
can be found in \cite{HJ},\cite{Seiberg}.
The important feature of (\ref{l2}) is that it describes
a classical conformal theory with the central charge
\beq
\label{c}
c={12 \over \gamma^2}~~~,
\eeq
which in our model is just the number of massless constituents
\beq
c=N~~~,
\eeq
see (\ref{gamma}). The latter fact is not surprising.
The massless constituents of our model are conformally
invariant in two dimensions. Under quantization
the conformal
symmetry acquires a central extension due to the conformal anomaly.
The central charge $c$ corresponds to this anomaly.

\section{Solutions to the Liouville theory}
\setcounter{equation}0

Equation (\ref{l2}) shows that the Liouville field is the
only dynamical variable of  induced Liouville
gravity\footnote{To avoid the confusion, let us note that
$\phi$ is the dynamical variable only in the classical theory,
in the quantum Liouville theory $\phi$ appears in the conformal
gauge but its contribution is compensated by the
contribution of corresponding ghosts.}.
By varying $\bar{I}_L$ with  fixed boundary
value of $\phi$, one finds the equation
\beq
\label{leq}
e^{-\gamma\phi}
\bar{\Delta}\phi=-{\mu\over 2\gamma}~~~.
\eeq
It follows from this equation that the physical metric
$g_{\mu\nu}=e^{\gamma \phi}\bar{g}_{\mu\nu}$ corresponds to a
space with  constant
negative curvature\footnote{It should be
noted that $R=-\mu/2$
results in further restrictions on the parameters of the
constituent fields of the Polyakov induced gravity. Namely,
the condition of the large masses, $M^2\gg R$, becomes
$M^2 \gg 48\pi |\lambda|/N$ where $M$ is a typical scale
for masses of the fields and $\lambda$ is given by Eq. (\ref{1.16}).
One can construct  models where this condition is satisfied.}
$R=-\mu/2$. This space is
locally a two-dimensional Lobachevsky space $H_2$. The corresponding
solution in the Lorentzian space-time is locally
anti-de Sitter ($AdS_2$).

There is also another approach to the variational problem.
One can start with the functional (\ref{ac1}) and consider
the Liouville field and metric $g_{\mu\nu}$ as independent
variables.
Then the equations of motion obtained from (\ref{ac1}) by
varying\footnote{The variational
procedure implies that the metric and $\phi$ are fixed on
the boundary.
The boundary term which depends on $n^\mu\sigma_{,\mu}$
was removed from
induced action (\ref{ac1})
in order to obey this
requirement.}
$\phi$ and $g_{\mu\nu}$ respectively, are:
\beq
R =
\gamma\Delta\phi~~~,
\label{eq: curvature}
\eeq
\beq
G_{\mu\nu}=0~~~,
\label{eq: dilaton1}
\eeq
\beq
G_{\mu\nu}\equiv\phi_{,\mu}\phi_{,\nu}-\frac
12g_{\mu\nu}\mid\nabla\phi\mid^2+
{2 \over \gamma}\left(
g_{\mu\nu}\Delta\phi-
\nabla_\mu\nabla_\nu\phi\right)
+{\mu \over 2\gamma^2} g_{\mu\nu}~~~.
\label{eq: dilaton1a}
\eeq
The trace of \req{eq: dilaton1} results in the relation
\beq
\Delta \phi = -{\mu \over 2\gamma}~~~
\label{*}
\eeq
which coincides with the Liouville equation (\ref{leq}). This justifies
considering the more general variational problem where
$\phi$ and the metric are independent
fields.

When $\phi$ obeys (\ref{leq})
\begin{equation}\label{stress}
G_{\mu\nu}=-4\pi\langle \hat{T}_{\mu\nu}\rangle ~~~,
\end{equation}
where $\hat{T}_{\mu\nu}$ is
the quantum stress-energy tensor of the constituents computed in the
considered approximation (\ref{eq: curvature}). Thus, (\ref{eq:
dilaton1}) is equivalent to relation (\ref{i1}) of the
Sakharov's induced
gravity. It should be noted that boundary conditions for $\phi$ which
are required to solve for it from (\ref{eq: curvature}) are related to the
choice of the quantum state \footnote{A detailed
discussion of this problem can be found
for instance in Ref. \cite{FIS}.}.

\bigskip

There is three types of solutions of the
Euclidean Liouville equations
(\ref{leq}): elliptic, parabolic and hyperbolic (see, e.g.
\cite{Seiberg}). These solutions correspond to different coordinate
maps on the Lobachevsky space $H_2$.

\bigskip

1. {\it Elliptic solutions.} The metric has the
form\footnote{We consider only the solutions which are
free from conical singularities}
\beq
\label{es1}
ds^2=
e^{\gamma\phi}d\bar{s}^2={16 \over \mu}{1 \over
(1-\rho^2)^2}d\bar{s}^2~~~,
\eeq
\beq
\label{fm1}
d\bar{s}^2=\rho^2 d\tau^2+d\rho^2~~~,~~~0\leq \tau \leq 2\pi~~~.
\eeq
We specify the flat metric by the boundary condition
\beq
\label{bcond}
0<\rho \leq \rho_+~~~.
\eeq
Solution (\ref{es1}) can be also written in the form
\beq
\label{metric}
ds^2=g(x)d\tau^2+{1 \over g(x)} dx^2~~~,
\eeq
\beq
\label{es2}
g(x)={\mu \over 4}x^2+\frac 12 x~~~,
\eeq
\beq
\label{ex}
x={8 \over \mu} {\rho^2 \over 1-\rho^2}~~~.
\eeq
This solution has the topology of a disk and
can be interpreted as a black hole instanton.
The horizon of the black hole is located at $x=0$ ($\rho=0$).
The normalization of the time coordinate $\tau$ is chosen
so that the
corresponding surface gravity constant is $g'(0)/2=1$.
One can make the surface gravity equal to another constant by
rescaling $\tau$.

The black hole solutions in the Lorentzian metric have the
anti-de Sitter geometry and we will call them $AdS_2$
black holes for the
brevity.

\bigskip

2. A {\it hyperbolic solution} depends on an integration
constant $m$ and has the form
\beq
\label{hs1}
ds^2=
e^{\gamma\phi}d\bar{s}^2={4 \over \mu}{m^2 \over
\rho^2 \sin^2(m\ln \rho)}d\bar{s}^2~~~.
\eeq
The flat metric $d\bar{s}^2$ is given by (\ref{fm1}). The hyperbolic
solution has the topology of a cylinder.  The flat space ${\cal M}$ is
defined by the boundary condition $\rho_-\leq \rho \leq \rho_+$ where
\beq
\label{bcondh}
e^{-\pi/m}< \rho_- < \rho_+ < 1~~~.
\eeq
The metric (\ref{hs1}) can be also written in the form (\ref{metric})
with
\beq
\label{hs2}
g={\mu \over 4} x^2+{4m^2 \over \mu}~~~,
\eeq
\beq
\label{hx}
x=-{4m \over \mu} \cot (m\ln\rho)~~~.
\eeq

\bigskip

3. {\it Parabolic solutions} can be obtained from
the hyperbolic ones in the limit $m\rightarrow 0$ and
look as follows
\beq
\label{ps1}
ds^2=
e^{\gamma\phi}d\bar{s}^2={4 \over \mu}{1 \over
(\rho \ln \rho)^2}d\bar{s}^2~~~,
\eeq
where $d\bar{s}^2$ is determined by (\ref{fm1}).
The coordinate $\rho$ subjects to the boundary condition
$\rho_-\leq \rho \leq \rho_+$, where $\rho_->0$ and $\rho_+<1$.
In the limit
when $\rho_- =0 $ the curvature of the solution has a
delta-function-like singularity.
The metric (\ref{ps1}) can be also written in the form (\ref{metric})
with
\beq
\label{ps2}
g={\mu \over 4} x^2~~~,
\eeq
\beq
\label{px}
x=-{4 \over \mu\ln\rho} ~~~.
\eeq
As follows from (\ref{ps2}) these solutions are analogous to
extremal black holes.

\bigskip

Note that Liouville solutions of all three types have the same
asymptotic behaviour at large $x$.  It is also worth pointing out that
each Liouville solution can be interpreted as a boundary theory of a
three-dimensional AdS-gravity and corresponds to a particular
three dimensional object
\cite{Martinec}
\footnote{It is interesting
to note that in a 4D gravity there also
exist 3 different families of Schwarzschild-(anti) de Sitter solutions.
These solutions are given by metric $ds^2=-A\, dt^2+dr^2/A+r^2 \,
d\Omega^2$, where $A=K-(2m/r)+\Lambda r^2/3$, $d\Omega^2=d\theta^2+
\mbox{sinn}^2\theta\,d\phi^2$. For an elliptic case,   $K=1$ and
$\mbox{sinn}\,\theta=\sin\theta$. For a hyperbolic case,
$K=-1$ and $\mbox{sinn}\,\theta=\sinh\theta$. For a parabolic case,
$K=0$ and $\mbox{sinn}\,\theta=\theta$. The elliptic solution (with
$K=1$) corresponds to a usual black hole. The hyperbolic solution
describes a black hole moving with a superluminar velocity, while the
hyperbolic solution describes a black hole moving with the velocity of
light.}.
Elliptic
solutions with conical singularities correspond to
massive particles in $AdS_3$, hyperbolic and
parabolic solutions may be related to non-extreme and extreme
three-dimensional black holes, respectively.
The non-extreme black holes correspond to a particular Liouville
theory
on a torus.

\section{Elliptic solutions}
\subsection{Thermodynamics of $AdS_2$ black holes}
\setcounter{equation}0

Let us now study black holes corresponding to elliptic solutions.
It is perhaps necessary at this stage to  explain
in what sense we expect constant curvature solutions
to behave like black holes.  The key assumption that we make is that the
Liouville field $\phi$ is an observable quantity, despite the fact that
it is not, strictly speaking, a fundamental field. It is rather introduced
for calculational convenience to make the effective action local
and defined formally by the relationship
$\phi = {1\over \gamma\Delta} R(g)$.
In effect, the observability of $\phi$ requires matter to
couple non-locally to the metric $g$ (i.e. to ${1\over \Delta} R$).
This does not occur at the classical level, but in the full quantum
effective action one does expect such terms to appear. It is on this
basis that we feel justified in assuming that $\phi$ is an observable
field, analogous to a dilaton.
Henceforth we treat the Liouville action as if it were a
dilaton gravity theory of the general form considered extensively in a
variety of references\cite{generic dilaton gravity}\footnote{
It is
worth noting that there exists at least one other theory, namely
 Jackiw-Teitelboim gravity, in which constant curvature black hole
solutions have been analyzed in some detail. In that case, the black
holes can be interpreted in terms of dimensionally reduced 2+1 Einstein
gravity (the so-called BTZ black holes\cite{BTZ}). Interestingly, the
Jackiw-Teitelboim theory was originally motivated by the connection
between constant curvature metrics and induced Liouville gravity. The
dilaton was considered to be a physically irrelevant lagrange
multiplier field needed to enforce the constant curvature equation.
Once the dilaton is taken seriously, it becomes clear that
Jackiw-Teitelboim gravity and induced Liouville gravity are quite
different theories, despite the fact that the solutions to both involve
constant curvature metrics.}.

\bigskip

We will interpret the induced Euclidean action (\ref{ac1})
considered on the elliptic solution (\ref{es1})
as $T$ times the free energy of the corresponding black hole. The black hole
is completely characterized by the temperature $T$ measured at the
boundary $x=x_+$
\beq
\label{temp}
T=(2\pi\sqrt{g(x_+)})^{-1/2}~~~.
\eeq
This condition defines $x_+$ in terms of $T$
\beq
\label{ex2}
x_+={1 \over \pi\mu T} \left(\sqrt{(4\pi T)^2+\mu}-4\pi T\right)~~~.
\eeq
The Liouville action on the elliptic solution is
\beq
\label{lac}
I_L(x_+)=-{\mu \over 2\gamma^2} x_+~~~.
\eeq
To calculate the induced action (\ref{ac1}) one also has to
know the value of the Polyakov action on the disk of the radius
$\rho_+$. The dependence of this action on $\rho_+$ can be found out
by making a scaling transformation (i.e. a conformal transformation
with
constant factor) to the disk of  unit radius
\beq
\label{disk}
N W[\bar{g}(\rho_+)]=
-{2 \over \gamma^2} \ln\rho_+ +C~~~.
\eeq
Here $C$ is a constant corresponding to the action on
the unit disk. It
does not depend on $T$ and can be omitted. From (\ref{ex}),
(\ref{lac})
and (\ref{disk}) we obtain for (\ref{ac1})
\beq
\label{eac1}
\Gamma=-{1
\over \gamma^2}\left({\mu \over 2}x_+ + \ln {\mu x_+ \over \mu x_+
+8}\right)~~~.
\eeq
This result can be immediately expressed in terms
of the temperature on the boundary by using (\ref{temp}) and
(\ref{ex2}).
By neglecting in (\ref{eac1}) a numerical constant we find
the free energy of the black hole
\beq
\label{fen}
F^{BH}(T)=\left. T\Gamma(x_+)\right|_{x_+=x_+(T)}
=-{1 \over \gamma^2}\left[
{1 \over 2\pi}\sqrt{(4\pi T)^2 +\mu}
+2T
\ln\left(\sqrt{(4\pi T)^2+\mu}-4\pi T\right)\right]
~~~.
\eeq
Thus, the black hole entropy defined by the standard relation,
is
\beq
\label{ent1}
S^{BH}(T)=-{\partial F^{BH} \over \partial T}={2 \over \gamma^2}
\ln\left(\sqrt{(4\pi T)^2+\mu}-4\pi T\right)~~~.
\eeq
(Here all numerical constants were omitted.)
The energy corresponding to this solution is
\beq
\label{engy}
E^{BH}(T)=-{1 \over 2\pi \gamma^2}\sqrt{(4\pi T)^2 +\mu}~~~.
\eeq
The variations of the energy and the entropy are related by
the first law
\beq
dE^{BH}=T dS^{BH}~~~.
\eeq
It should be noted that the given thermodynamical system is not
stable because its heat capacity is negative for all
values of T
\beq
\label{cap}
c(T)=T{\partial S^{BH} \over \partial T}=-{1 \over \gamma^2}
{8\pi T \over \sqrt{(4\pi T)^2 +\mu}}~~~.
\eeq
The energy of this black hole decreases when its temperature grows.
Increasing the temperature corresponds to moving the boundary
closer to the black hole horizon ($x_+\rightarrow 0$).

As follows from (\ref{cap}) the heat capacity is increasing at small
temperatures and one may speculate that in this region
the black hole may be in a quasi-equilibrium state. Thus at
low temperatures our description of the black hole in terms
of a canonical ensemble may be justified. On the other hand, at high
temperatures $T$ the system is very unstable and one should expect
a phase transition. Other possible phases of the theory may be related
to other solutions of ILG.

\newpage

\subsection{Black hole statistical mechanics}

\noindent
{\bf A. Black hole canonical ensemble in terms of the constituents}
\bigskip

\noindent

We now show how the black hole can be
described in terms of statistical mechanics of the constituents
of ILG.
To this aim let us consider the constituents which propagate
in the static region of the black hole, outside the horizon,
and compare the canonical ensembles of the black hole and
the constituents.

It is well known that the description of statistical-mechanics
in the presence of a Killing horizon meets difficulties
due to the divergences of the density of states near the horizon
\cite{FF:98a}.
The standard method to proceed in this situation is to
introduce a cutoff near the horizon at some proper distance $\epsilon$.
This cutoff can be considered as an inner boundary.
Then the regularized free energy $F(\epsilon,T)$ of the
constituents has the standard definition
\beq
\label{fedef}
e^{-F/T} =\mbox{Tr} e^{-(\hat{H}-{\cal E}_v)/T}~~~,
\eeq
where $T$ is the temperature of the system. The operator $\hat{H}$
is normally ordered total Hamiltonian of all the constituents,
and ${\cal E}_v$ is the zero-point energy. In our approach
the massive constituents are very heavy and contribute only
to the cosmological constant. It means that in the considered
approximation all the effect of these fields is in the vacuum energy
${\cal E}_v$. Thus, one can rewrite (\ref{fedef}) as
\beq
\label{fedef2}
e^{-F/T} =
e^{{\cal E}_v/T}
\left(\mbox{Tr} e^{-\hat{H}_0/T}\right)^N=
e^{({\cal E}_v-NF_0)/T}
~~~,
\eeq
where $\hat{H}_0$ and $F_0$ are, respectively, the normally ordered
Hamiltonian and the statistical-mechanical
free energy of a single massless constituent.

It is also instructive to represent the free energy (\ref{fedef})
of the constituents
in another form, in terms of the induced Euclidean action
$\Gamma(x_+,\epsilon,T)$. The Euclidean theory is formulated
on the elliptic solution (\ref{es1}) with an inner boundary at
$\rho_-$ determined by the cutoff,
\beq
\rho_-\simeq e^{-\gamma\phi_-/2} \epsilon~~~,
\eeq
where $\phi_-$ is the value of the Liouville field
at this boundary ($\phi_-$ is a finite constant at
$\epsilon\rightarrow 0$). The both functionals are related as
\beq
\label{acfe}
\Gamma(x_+,\epsilon, T)+C=F(\epsilon, T)T^{-1}~~~,
\eeq
where
$C$ is a possible finite numerical constant
which accounts for, according to \cite{Allen}, the difference
between field-theoretical and statistical-mechanical computations.
$x_+$ in (\ref{acfe}) is expressed in terms of $T$ by (\ref{ex2}).
The induced action is determined by (\ref{ac1}),
\beq
\label{acc}
\Gamma(x_+,\epsilon, T)=I_L(x_+,\epsilon, T)+NW[K_\epsilon]~~~.
\eeq
The action $W[K_\epsilon]$ is the Polyakov action on the annulus
$K_\epsilon$
\beq
ds^2=\rho^2 d\tau^2+d\rho^2~~~,~~~\rho_-(\epsilon)
\leq \rho \leq \rho_+~~~.
\eeq
It is related to the Polyakov action on the cylinder $Q_b$
\beq
\label{ult1}
ds^2=d\tau^2+dy^2~~~,~~~0\leq \tau\leq 2\pi~~~,
\eeq
\beq
\label{ult2}
dy=d\ln \rho~~~,~~~0\leq y \leq b~~~,
\eeq
by the conformal transformation,
\beq
W[K_\epsilon]=W[Q_b]-{b \over 12}~~~,
\eeq
\beq
\label{cut}
b=\ln {\rho_+ \over \rho_-}\simeq \ln {x_+ T \over \epsilon}~~~.
\eeq
Because of the conformal invariance the spectrum of
single particle excitations of
two-dimensional massless scalars coincides
with the spectrum of these fields
on related flat ultrastatic space (\ref{ult1})
(see for the details,
e.g.,
\cite{FF:98a})). As a result of this property, the effective action on
the cylinder is related to the statistical-mechanical
free energy of a single massless constituent
in a simple way
\beq
\label{cacfe}
T W[Q_b]=F_0(\epsilon,T)~~~.
\eeq
Equations (\ref{fedef2}), (\ref{acfe}), (\ref{acc}) and (\ref{cacfe})
enable us to find the expression for the vacuum energy
\beq
{\cal E}_v(\epsilon, T)=T\left(I_L(x_+,T)-{N \over 12}
\ln{\rho_+ \over \rho_- }+C\right)~~~.
\eeq
(Here we took into account that the Liouville action
on the annulus and that on the disc differ by a constant in the limit
$\epsilon\rightarrow 0$.)
The divergence of the vacuum energy is the result of
the divergence of the density of states near the horizon.

The statistical-mechanical entropy of a two-dimensional gas
on the interval with the size $b$
can be computed exactly in the
limit of large $b$, see, e.g. \cite{FFZ:96}.
If the temperature is $\beta^{-1}$, the free energy, entropy and energy
are
\beq
\label{fe}
F^{\mbox{sm}}(b,\beta)
\simeq -{\pi \over 6} {b \over \beta^2} -
{1 \over 2\beta} \ln{\beta
\over 2b}~~~,
\eeq
\beq
\label{sment}
S^{\mbox{sm}}(b,\beta)
\simeq {\pi \over 3} {b \over \beta} +\frac 12 \ln{\beta
\over 2b} ~~~,
\eeq
\beq
\label{en}
E^{\mbox{sm}}(b,\beta)
\simeq {\pi \over 6} {b \over \beta^2} -{1 \over 2\beta}
~~~,
\eeq
where all constants which are finite at large $b$ are omitted.

The parameter $\beta$ coincides with the periodicity of the
Euclidean time $\tau$ in (\ref{ult1}). In our case $\beta=2\pi$.
By taking into account that the physical temperature is $T$
one can write
\beq
F_0(\epsilon, T)=2\pi T F^{\mbox{sm}}(b,2\pi)~~~.
\eeq
This relation in combination with (\ref{fedef2}),
(\ref{fe})--(\ref{en}) gives the
following result
\beq
\label{1}
F(\epsilon,T)=
E(\epsilon,T)-
TS(\epsilon,T)~~~,
\eeq
\beq
\label{2}
E(\epsilon,T)={\cal E}_v(\epsilon,T)+2\pi N T E^{\mbox{sm}}(b,2\pi)
=TI_L(x_+)+C T~~~,
\eeq
\beq
\label{3}
S(\epsilon,T)=NS^{\mbox{sm}}(b,2\pi)~~~.
\eeq
Note that
the total energy $E(\epsilon,T)$
is finite in the limit $\epsilon\rightarrow 0$
because the divergence of
zero-point fluctuations is compensated by the divergence of the thermal
excitations of the massless constituents.  Moreover, by using
(\ref{ex2}) and (\ref{lac}) we find that
\beq
E^{BH}(T)=E(0,T)+CT~~~,
\eeq
where $C$ is a numerical constant. By using an arbitrariness
in relation (\ref{acfe}) between the Euclidean action and the
free energy one can always make $C$ equal to 0.
After this "normalization" the statistical-mechanical
energy of the induced gravity constituents
coincide with black hole
energy  (\ref{engy}).

Let us consider now the entropy of the constituents.
Because this quantity corresponds to the fields propagating
outside the horizon it can be interpreted as the entanglement
entropy. As is well known, the entanglement entropy
is divergent when the cutoff is removed. This is its
key distinction
from the black hole entropy $S^{BH}$. However,
as follows from (\ref{ent1}), (\ref{sment}) and (\ref{3}),
the two entropies
are related\footnote{An analogous
subtraction formula for
two-dimensional black holes was discussed in \cite{FFZ:96}.
However,
that paper was dealing with quantum corrections to the
black hole entropy rather than to the entropy itself.},
\beq
\label{r2}
S^{BH}(T)=
S(\epsilon, T)-
S_R(\epsilon)~~~,
\eeq
\beq
\label{entrin}
S_R(\epsilon)
\simeq -{2 \over \gamma^2}
\ln\epsilon
-{6\over \gamma^2}\ln|\ln\epsilon|~~~.
\eeq
It is not difficult to see that $S_R(\epsilon)$
can be interpreted as
the entropy of the constituents on the Rindler space
with the same cutoff
\beq
\label{rind}
ds^2=\rho^2 d\tau^2+d\rho^2~~,~~\epsilon< \rho \leq 1~~,
~~0\leq \tau \leq 2\pi~~~.
\eeq
It is important that if one considers variations
of the parameters of the black hole at fixed value of the
parameter $\epsilon$, changes of entropy (\ref{3})
coincide with changes of thermodynamical entropy (\ref{ent1})
of a black hole
\beq
\label{r1}
\Delta S^{BH}(T)=\Delta S(\epsilon, T)~~~.
\eeq
Thus, from the point of view of thermodynamics the two entropies
are equivalent. Moreover, the above relation
does not depend on the choice of the regularization prescription.
Instead of using the cutoff near the horizon one can arrive at
(\ref{r1}) using the dimensional or Pauli-Villars regularization
schemes which also enable one to eliminate the
divergences related to the horizon (see for details, e. g.
\cite{FF:98a}).

\bigskip

The analysis of this Section demonstrates that the thermodynamics of
$AdS_2$ black hole has a statistical-mechanical explanation in
terms of the constituents of ILG. One can conclude that the
constituents are the real degrees of freedom of the black
hole. It is interesting to point out the black hole entropy is related
to the massless constituents only. It does not mean, however, that the
massive fields are irrelevant. As we saw, these constituents
provide the finite cosmological constant and give a contribution
to the vacuum energy which depends on the black hole parameters.

\bigskip

\noindent
{\bf B. Conformal field theory}
\bigskip

The above result for the entropy can be computed by means of
a conformal field theory (CFT) along the lines of
computations of the entropy of BTZ black holes \cite{Strominger}.
Massless constituents of ILG are described
by a CFT with
the central charge $c=N$, see (\ref{c}).

The computation of the entropy is as follows \cite{F:98b},\cite{HLW}.
The relation between the
Hamiltonian of the system and generators of the Virasoro algebra follow
from the representation of the metric (\ref{ult1}) in the form
\begin{equation}\label{ult3}
d\bar{s}^2=\left({{ b} \over \pi}\right)^2(-d\eta^2+dz^2)=
\left({ { b} \over \pi}\right)^2 dudv~~~,
\end{equation}
\begin{equation}\label{uv}
u={z+\eta \over 2}~~~,~~~v={z-\eta \over 2}~~~.
\end{equation}
Consequently,
\begin{equation}\label{der}
\partial_t={\pi \over 2{ b}}(\partial_u-\partial_v)~~~.
\end{equation}
In (\ref{ult3}) the coordinate $z$ ranges from $0$ to $\pi$.
This corresponds to a theory on an interval where the points $z=0$ and $z=\pi$
are independent.
In order to carry out the computations it is convenient
to pass to a theory where $z$ is a periodic coordinate.
This can be done if one considers two equivalent CFT's on the
intervals with the length $\pi$ and makes from
them a CFT on a circle by gluing together the ends of the
intervals.  In the obtained theory
$z$ has the periodicity $2\pi$.

One has two copies of the Virasoro algebra where
the elements $L_n$ and $\bar{L}_n$ can be defined in a standard way
as the generators of the coordinate transformations,
$\delta u=e^{inu}$ and
$\delta v= e^{inv}$, respectively.
As a result of relation (\ref{der}),
the Hamiltonian of massless constituents which generates transformations
along the Killing time $t$ coincides with the operator
$\pi(L_0-\bar{L}_0)/2b$.
Similarly, translations of the
system along $y$ are generated by the momentum
$\pi(L_0+\bar{L}_0)/2b$.
Because the system is at rest the average momentum
is zero. On the other hand, the average value of
the energy is $NE^{\mbox{sm}}(b,\beta)$.
This fixes the average values $h$ and $\bar{h}$
of $L_0$ and $\bar{L}_0$, respectively.
In the leading approximation
\begin{equation}\label{levels}
h=-\bar{h}={N \over 6}{{ b}^2 \over \beta^2}~~~.
\end{equation}
In the limit when $ b$ is large ($\epsilon$ goes
to zero), $h \gg 1$ and one can use Cardy's formula
to estimate the degeneracy of $L_0$ and $\bar{L}_0$.
In this approximation the total degeneracy $D$ is
\begin{equation}\label{deg1}
\ln D=2\pi\sqrt{{ch \over 6}}+2\pi\sqrt{{c|\bar{h}| \over 6}}
\end{equation}
and by taking into account that in our case the central charge
$c=N$ we find
\begin{equation}\label{deg2}
\ln D=2N {\pi \over 3}{ b \over \beta}~~~.
\end{equation}
Finally, we have to remember that $D$ is the number of states
of the system with the doubled Hilbert space which results
from the trick of imposing periodicity of the coordinate $z$.
The real number
of states of the
system we are interested in is $\sqrt{D}$.
Thus, the entropy is
\begin{equation}\label{cftent}
\frac 12 \ln D=NS^{\mbox{sm}}
\end{equation}
and it coincides exactly with the required value
in Eq. (\ref{sment}) for $N$ fields.

A remark concerning fixing of the Virasoro level is in order.
In our calculation the eigenvalues
$h$, $\bar{h}$ are not connected with the
observable energy (\ref{engy}) of a
black hole but are determined by the average value of the
normally ordered Hamiltonian, i.e., by the thermal energy
of the fields.

\section{Hyperbolic and parabolic solutions}
\setcounter{equation}0

It is also worth studying thermodynamics of
parabolic and hyperbolic instantons of ILG.
Below we briefly comment on its features.

The solutions have two boundaries, and the fiducial
space $\bar{\cal M}$ in
the both cases (see (\ref{hs1}) and (\ref{ps1}))
is an  annulus.
For this reason there are no difficulties related to the
horizon as in the case of $AdS_2$ black holes.  This
automatically provides the agreement between thermodynamics
of the solutions and statistical mechanics of the constituents.
To begin with it is convenient to replace the fiducial
space $\bar{\cal M}$ in (\ref{ac1}) to
a cylinder by making  coordinate and conformal transformations and
changing the Liouville field.
Then (\ref{hs1}) and (\ref{ps1}) can be
written as
\beq
\label{hps1}
ds^2=e^{\gamma\phi}d\bar{s}^2~~~,
\eeq
\beq
\label{cm}
d\bar{s}^2=d\tau^2+dy^2~~~,
\eeq
\beq
0\leq \tau \leq \beta~~~,~~~y_-\leq y\leq y_+~~~.
\eeq
In case of the hyperbolic solution
\beq
\label{hs3}
e^{\gamma\phi}={4 \over \mu}{1 \over \sin^2y}
\eeq
and $\beta=2\pi m$ where $m$ is defined in (\ref{hs1}).
The boundary coordinates subject to
the restrictions $y_->0$ and $y_+<\pi$.
For the parabolic solution
\beq
\label{ps3}
e^{\gamma\phi}={4 \over \mu}{1 \over y^2}
\eeq
where $y_+<0$. In this case $\beta$ is an arbitrary parameter.
In fact, the parabolic metric does not change when one rescales
the coordinates $y$ and $\tau$ with the same coefficient. Thus
one can choose any  periodicity for $\tau$ by appropriate
redefinition of the boundary values $y_\pm$.

The induced action (\ref{ac1}) for hyperbolic and parabolic solutions
has a very simple form because the Polyakov action on the cylinder
(\ref{cm}) is determined by the statistical-mechanical
free energy (\ref{fe}) at the
temperature $\beta^{-1}$ on the interval of the length $b=y_+-y_-$
\beq
\label{polfe}
W(y_+-y_-,\beta)=
\beta F^{\mbox{sm}}(y_+-y_-,\beta)~~~.
\eeq
Note, as we explained in the previous section,
$F^{\mbox{sm}}(y_+-y_-,\beta)$ coincides with
the free energy of a single massless scalar constituent of ILG.
One can represent the induced action (\ref{ac1})
in the form
\beq
\label{iac}
\Gamma(y_+,y_-,\beta)=\beta \left[f(y_+)-f(y_-)+
NF^{\mbox{sm}}(y_+-y_-,\beta)\right]~~~.
\eeq
The function $f(y)$ can be easily found by calculating the
Liouville action (\ref{l2}) for the given solutions.
One has
\beq
\label{hf}
f(y)={1 \over \pi \gamma^2}\left( 4 \cot y +2y-{\gamma \over 2}
\phi(y)\right)~~~
\eeq
for the hyperbolic solution, and
\beq
\label{pf}
f(y)={1 \over \pi \gamma^2}\left({4 \over y} -{\gamma \over 2}
\phi(y)\right)~~~
\eeq
for the parabolic one.
As follows from these expressions, the parabolic action is a
limiting form of the hyperbolic functional at the small
values of $y_{\pm}$.

The thermodynamics of these solutions is the thermodynamics of
a system in a finite volume. The corresponding
thermodynamical state is
fixed by the boundary values $y_{\pm}$ which characterize the size
of the system and the temperature $T$ measured at one of the
boundaries.
The free energy is determined by the induced action (\ref{iac})
as usually,
$F(T,y_+,y_-)=T\Gamma(\beta,y_+,y_-)$.
The inverse temperature $T^{-1}$ is the
circumference length of the boundary
and is proportional to
$\beta$.  Thus, one immediately concludes from (\ref{iac})
that the thermodynamical
entropy  coincides with the statistical entropy of
the massless constituents of ILG
\beq
\label{eq3}
S(T,y_+,y_-)=NS^{\mbox{sm}}(\beta,y_+-y_-)~~~,
\eeq
where
\beq
S(T,y_+,y_-)=-\left(
{\partial F(T,y_+,y_-) \over \partial T}
\right)_{y_+,y_-}~~~,
\eeq
and $S^{\mbox{sm}}$ is defined
in (\ref{sment}).
The equality (\ref{eq3}) is more strong than relation
(\ref{r1}) for the elliptic solutions.

To conclude this Section we should note that the
first law for the hyperbolic and the parabolic solutions
has a more general form
\beq
\label{hp1law}
\delta E=T\delta S-p_+\delta y_+ -p_-\delta y_-~~~,
\eeq
where
\beq
E=F+TS
\eeq
is the energy of the system and  the quantities
\beq
p_\pm=-
\left(
{\partial F(T,y_+,y_-) \over \partial y_\pm}
\right)_{T,y_\mp}~~~
\eeq
can be can be interpreted as pressures at the boundaries.
They can be computed with the help of (\ref{iac})--(\ref{pf}).

\section{ILG dual to BTZ black holes}
\setcounter{equation}0

Our discussion includes as a particular example
the induced Liouville gravity which is dual to the
BTZ black hole sector of the three-dimensional gravity
with the negative cosmological constant ($AdS_3$ gravity).
The Euclidean action of this theory
\cite{BM}
\beq
\label{ads}
I^{(3)}=-{1 \over 16\pi G}\left[\int d^3x\sqrt{g} (R+{2 \over l^2})
+\int_{\infty} d^2x\sqrt{h} K\right]~~~,
\eeq
where $K$ is the extrinsic curvature of the asymptotic boundary,
corresponds to a canonical free energy of the system.
Let us consider for simplicity a static BTZ black hole with the
mass $M$. Functional (\ref{ads})
taken on the corresponding instanton has the canonical form
\beq
\label{btz}
I^{(3)}=\beta M^{BTZ} - S^{BTZ}~~~,
\eeq
where $\beta$ and $S^{BTZ}$ are the inverse temperature and the entropy
of the black hole, respectively.
The parameters of the black hole are determined by
the radius $r_+$ of the horizon
\beq
\label{mass}
M^{BTZ}={r_+ \over 8G l^2}~~~,
\eeq
\beq
\label{tempr}
\beta={2\pi l^2 \over r_+}~~~,
\eeq
\beq
\label{ent}
S^{BTZ}={2\pi r_+ \over 4G}~~~.
\eeq
It is instructive to represent the free energy, entropy and mass
of the black hole in terms of the inverse temperature
\beq
\label{btzfe}
F^{BTZ}(l,\beta)=\beta^{-1} I^{(3)}=-{\pi^2 l^2 \over 2G \beta^2}
\eeq
\beq
\label{btzent}
S^{BTZ}(l,\beta)={\pi^2 l^2 \over G \beta}~~~,
\eeq
\beq
\label{btzmass}
M^{BTZ}(l,\beta)={\pi^2 l^2 \over 2G \beta^2}~~~.
\eeq
Let us now demonstrate that this black hole is thermodynamically
equivalent to a certain type of ILG. We begin with Euclidean ILG
with zero cosmological constant ($\mu=0$) which is a particular
case of ILG models. The boundary of a Euclidean BTZ black hole
is a torus, see, e.g., \cite{BBO}. This leads us to
consider ILG where the background ${\cal M}$ has
the topology of a torus.
The induced action of this theory is (see (\ref{ac1}))
\beq
\label{tlac}
\Gamma[g]= \bar{I}_L[\phi]+ NW[\bar{\cal M}]~~~,
\eeq
\beq
\label{tl2}
\bar{I}_L[\phi]=
-{1 \over 2\pi}\int_{\bar{\cal M}}
d^2y(\bar{\nabla}\phi)^2~~~.
\eeq
The flat space $\bar{\cal M}$ is a torus
\beq
\label{torus}
d\bar{s}^2=d\tau^2+dx^2~~~.
\eeq
We assume that
\beq
0< \tau \leq \beta~~~,~~~0< x \leq b~~~.
\eeq
The solutions to this theory are constant $\phi$
and on the solutions
the induced action coincides
with $NW[\bar{\cal M}]$.
Thus, the on-shell Euclidean action of ILG corresponds to the
free energy of massless gas on a circle of  length $b$ at the
temperature $\beta^{-1}$.
In the limit when  size $b$ is large one can
find the action on the torus from the leading order expression
(\ref{fe}) for the free energy on the interval of  size $b$.
We thus obtain
\beq
\label{tfe2}
F(b,\beta)\simeq -N{\pi \over 6} {b \over \beta^2}~~~.
\eeq
This free energy corresponds to the CFT with the integer
central charge $c=N$. Consider now the BTZ black hole
with the same central charge $c$ \footnote{The conformal
Virasoro algebra with an integer central charge has a number of
interesting features which were discussed in \cite{B}.}
\beq
\label{cc}
c_{BTZ}={3l \over 2G}=N~~~.
\eeq
This charge corresponds to the
group of diffeomorphisms at the asymptotic infinity of the BTZ
black hole.  By now replacing  $N$ in (\ref{tfe2}) by $c_{BTZ}$
and putting $b=2\pi l$ we come to the identity
between the ILG and BTZ free energy
(\ref{btzfe})
\beq
\label{btzcft}
\left. F^{BTZ}(l,\beta)=F(2\pi l,\beta)\right|_{N=c_{BTZ}}~~~.
\eeq
The corresponding identities can be established for the
entropy (\ref{btzent}) and energy (\ref{btzmass})
of the black hole.
Therefore, the static
BTZ black hole is {\it thermodynamically} equivalent to $1+1$
ILG having $c_{BTZ}$ massless fields and given on the circle of
the radius $l$ ($l$ is related to the curvature radius of the
corresponding $AdS_3$ geometry, see (\ref{ads})).

\bigskip

This result can be also generalized to rotating BTZ black holes.
If the angular velocity $\Omega$ of the black hole is not zero, the
free energy (\ref{btzfe}) is replaced to \cite{BBO}
\beq
\label{rotbtz}
F^{BTZ}(\Omega,l,\beta)=
-{\pi^2 l^2 \over 2G\beta^2 (1-l^2\Omega^2)}~~~.
\eeq
The corresponding two dimensional system equivalent
to this black hole is a rotating quantum gas.
To be more specific, consider a massless gas
where  half of the quanta
are rotating clock-wise with  angular velocity $\Omega$ and
the other half are moving in the opposite direction.
The corresponding free-energies of these quanta denoted by
$F_{\pm}$  are determined by the "boosted" partition functions
\beq
\label{rotfe}
e^{-\beta F_{\pm}}=\mbox{Tr} e^{-\beta (\hat{H}\pm \Omega \hat{M})}~~~,
\eeq
where $\hat{H}$ is the Hamiltonian and $\hat{M}$ is the angular
momentum related to the momentum $\hat{P}$ along the circle as
\beq
\hat{M}={b \over 2\pi} \hat{P}~~~.
\eeq
($b/(2\pi)$ is the radius of the circle). The energy
of a relativistic massless particle equals
the modulus of its momentum. Therefore
\beq
\label{rotfepm}
e^{-\beta F_{\pm}}=\mbox{Tr} e^{-\beta(1\pm \bar{\Omega})\hat{H}}~~~,
\eeq
\beq
\label{om}
\bar{\Omega}={b \Omega \over 2\pi}~~~.
\eeq
Thus, when $b$ is large one can find $F_{\pm}$ with the
help of (\ref{fe})
\beq
\label{tfe3}
F_{\pm}(\Omega,b,\beta)\simeq -{\pi b \over 6\beta^2(1\pm
\bar{\Omega})}~~~.
\eeq
The total free-energy of the system is
\footnote{A similar computation can be found
in \cite{HHTR}.}
\beq
\label{rotfe2}
F(\Omega,b,\beta)={N \over 2} F_{+}(\Omega,b,\beta)+
{N \over 2} F_{-}(\Omega,b,\beta)\simeq
-N{\pi \over 6} {b \over \beta^2(1-\bar{\Omega}^2)}
~~~.
\eeq
Now by taking into account (\ref{om}) and that $b=2\pi l$
we can generalize the relation (\ref{btzcft})
\beq
\label{btzcft2}
\left. F^{BTZ}(\Omega,l,\beta)=
F(\Omega,2\pi l,\beta)\right|_{N=c_{BTZ}}~~~.
\eeq
Once one has the  thermodynamic relation (\ref{btzcft2}) between the
BTZ black hole and a two-dimensional rotating gas, the
statistical-mechanical explanation of the two-dimensional entropy
can be considered as an explanation of the black hole
entropy \cite{Strominger}.
This, however, cannot be taken as a satisfactory explanation
of the entropy, because the degrees of freedom of the
two-dimensional theory have nothing to do with the
degrees of freedom of the black hole.

\section{Discussion}

Recent interesting computations of the entropy
of extremal \cite{Peet},\cite{StVa} and
BTZ \cite{Strominger} black holes leave open the
essential question about the real degrees of freedom
of the black hole. The difficulty is that
the computations
concern not black holes themselves but dual
systems in flat space-times.
Some help in resolving this difficulty may come from
studying models of Sakharov's induced gravity
\cite{FFZ:97}--\cite{FF:98b}.  In this theory the degrees
of freedom  responsible for the black hole entropy
are the constituents which induce the Einstein gravity in the
low-energy limit.

The Liouville induced gravity considered in
this paper is a suitable "firing range" to study this issue.
We have shown that the entropy of induced $AdS_2$
black holes is equivalent to the entanglement entropy
of the massless constituents. In two dimensions
the divergence usually encountered in
the definition of the entanglement entropy does
not depend on the thermodynamical
parameters of the black hole and, hence, is not observable in
physical processes.
Therefore, going to two-dimensional induced gravity
leads to an essential simplification.
In higher-dimensional models the relation between the
entanglement and black hole entropies is more complicated.
It always requires subtraction a non-trivial Noether charge
related to non-minimal couplings of the constituents \cite{FF:97a}.
Another distinction is that in higher dimensions
the main contribution to the entanglement entropy is
determined only by the constituents with the Planck mass localized
in the vicinity of the horizon. In ILG the entropy
is related to the fields which are localized in the entire
black hole exterior.

The induced Liouville gravity has solutions of different types
which may find different applications. The important example
are the solutions with zero cosmological constant on a
torus. On the level of thermodynamics they are equivalent
to BTZ black holes. It would be interesting to see
whether there is a similar relation of the
found solutions, including $AdS_2$ black holes, and
thermodynamics of the objects in the three-dimensional $AdS$ space.
Another interesting problem is to analyze along the lines of
Section 6 the theory
\cite{Carlip:98b}, \cite{Solodukhin:98}
which appears in the near horizon limit of generic black holes.
We leave these issues for future publication.

\noindent
\section*{Acknowledgments}

\indent This work was
partially supported  by the Natural
Sciences and Engineering
Research Council of Canada.


\newpage


\newpage

\begin{thebibliography}{9}

\bibitem{Bekenstein:72}  J.D. Bekenstein,
Nuov. Cim. Lett. {\bf 4}, 737 (1972).

\bibitem{Hawking:75} S.W. Hawking,
Comm. Math. Phys.
{\bf 43}, 199 (1975).

\bibitem{FF:98a} V. Frolov and D. Fursaev,
Class. Quantum Grav. {\bf 15}, 2041 (1998).

\bibitem{Peet} A.W. Peet, Class. Quantum Grav. {\bf 15}, 3291 (1998).

\bibitem{StVa} A. Strominger and C. Vafa, Phys. Lett. {\bf B379},
99 (1996)

\bibitem{Strominger} A. Strominger, JHEP {\bf 02}, 009 (1998).

\bibitem{BTZ} M. Ba\~nados, C. Teitelboim, and J. Zanelli,
Phys. Rev. Lett. {\bf 69}, 1849 (1992).

\bibitem{BH} J.D. Brown and M. Henneaux, Comm. Math. Phys.
{\bf 104}, 207 (1986).

\bibitem{CHD} O. Coussaert, M. Henneaux, and P. van Driel,
Class. Quantum Grav. {\bf 12}, 2961 (1995).

\bibitem{Carlip:98a} S. Carlip, Class. Quantum Grav. {\bf 15},
3609 (1998).

\bibitem{Carlip:95} S. Carlip, Phys. Rev. {\bf D51}, 631 (1995).

\bibitem{Carlip:98b} S. Carlip, {\it Black Hole Entropy from
Conformal Field Theory in Any Dimension}, e-Print Archive:
hep-th/9812013.


\bibitem{Solodukhin:98} S.N. Solodukhin, {\it Conformal Description
of Horizon's States}, e-Print Archive: hep-th/9812056.


\bibitem{Martinec} E. Martinec, {\it Conformal Field Theory,
Geometry and Entropy}, e-Print Archive: hep-th/9809021.

\bibitem{St:98} A. Strominger, {\it $AdS_2$ Quantum gravity
and String Theory}, e-Print Archive: hep-th/9809027.


\bibitem{generic dilaton gravity}  J. Gegenberg, G. Kunstatter,
and D. Louis-Martinez, Phys. Rev.  {\bf D51}, 1781 (1995); {\it ibid} Phys.
Letts. {\bf B321}, 193 (1994).  %

\bibitem{Jacobson:94} T. Jacobson,
{\it Black Hole Entropy in Induced Gravity},
e-Print Archive: gr-qc/9404039.

\bibitem{FFZ:97} V.P. Frolov, D.V. Fursaev,
and A.I. Zelnikov, Nucl. Phys. {\bf B486}, 339 (1997).

\bibitem{FF:97a} V.P. Frolov, and D.V. Fursaev,
Phys. Rev. {\bf D56}, 2212 (1997).

\bibitem{FF:98b} V. Frolov and D. Fursaev,
Phys. Rev. {\bf D58}, 124009 (1998).


\bibitem{Sakharov:68} A.D. Sakharov,
Sov. Phys. Doklady, {\bf 12} (1968) 1040.


\bibitem{Polyakov} A.M. Polyakov, Phys. Lett. {\bf 103B}, 207 (1981).

\bibitem{HJ} E. D'Hoker and R. Jackiw, Phys. Rev. {\bf D26}, 3517
(1982).

\bibitem{Seiberg} N. Seiberg, Progress of Theoretical Physics Suppl.
{\bf 102}, 319 (1990).


\bibitem{FIS} V.P. Frolov, W. Israel, and S.N. Solodukhin,
Phys. Rev. {\bf D54} (1996) 2732.

\bibitem{FFZ:96} V.P. Frolov, D.V. Fursaev, and A.I. Zelnikov,
Phys. Rev. {\bf D54}, 2711 (1996).

\bibitem{Allen} B. Allen, Phys. Rev. {\bf D33}, 3640 (1986).


\bibitem{F:98b} D.V. Fursaev, {\it A Note on Entanglement
Entropy and Conformal Field Theory},
e-Print Archive: hep-th/9811122.

\bibitem{HLW} C. Holzhey, F. Larsen, and F. Wilczek,
Nucl. Phys. {\bf B424}, 443 (1994).

\bibitem{BM} M. Ba\~nados and F. Mendez, Phys. Rev. {\bf D58},
104014 (1998), e-Print Archive: hep-th/9806065.

\bibitem{BBO} M. Ba\~nados, T. Brotz, and M. Ortiz,
{\it Boundary Dynamics and the Statistical Mechanics of the
$2+1$ Dimensional Black Hole},
e-Print Archive: hep-th/9802076.

\bibitem{B} M. Ba\~nados, {\it Embeddings of the Virasoro
Algebra and Black Hole Entropy},
e-Print Archive: hep-th/9811162.

\bibitem{HHTR} S.W. Hawking, C.J. Hunter, and M.M. Taylor-Robinson,
{\it Rotation and AdS/CFT Correspondence},
e-Print Archive: hep-th/9811056.

\end{thebibliography}
\end{document}